\documentclass[ApJL, twocolumn]{aastex631}

\usepackage{graphicx}
\usepackage{amsmath}
\usepackage{color}
\usepackage[normalem]{ulem}

\newcommand{\SPA}{School of Physics and Astronomy, Monash University, Clayton VIC 3800, Australia}
\newcommand{\OzGravMonash}{OzGrav: The ARC Centre of Excellence for Gravitational Wave Discovery, Clayton VIC 3800, Australia}

\shorttitle{Which black hole is spinning?}
\shortauthors{Adamcewicz et al.}

\begin{document}

\title{
Which black hole is spinning?\\
Probing the origin of black-hole spin with gravitational waves}

\author{Christian Adamcewicz}
\email{christian.adamcewicz@monash.edu}
\affiliation{\SPA}
\affiliation{\OzGravMonash}

\author{Shanika Galaudage}
\affiliation{Université Côte d'Azur, Observatoire de la Côte d'Azur, CNRS, Laboratoire Lagrange, Bd de l'Observatoire, F-06304 Nice, France}
\affiliation{Université Côte d'Azur, Observatoire de la Côte d'Azur, CNRS, Artemis, Bd de l'Observatoire, F-06304 Nice, France}

\author{Paul D. Lasky}
\affiliation{\SPA}
\affiliation{\OzGravMonash}

\author{Eric Thrane}
\affiliation{\SPA}
\affiliation{\OzGravMonash}

\begin{abstract}
Theoretical studies of angular momentum transport suggest that isolated stellar-mass black holes are born with negligible dimensionless spin magnitudes $\chi \lesssim 0.01$.
However, recent gravitational-wave observations indicate $\gtrsim 40\%$ of binary black hole systems contain at least one black hole with a non-negligible spin magnitude.
One explanation is that the first-born black hole spins up the stellar core of what will become the second-born black hole through tidal interactions.
Typically, the second-born black hole is the ``secondary'' (less-massive) black hole, though, it may become the ``primary'' (more-massive) black hole through a process known as mass-ratio reversal.
We investigate this hypothesis by analysing data from the third gravitational-wave transient catalog (GWTC-3) using a ``single-spin'' framework in which only one black hole may spin in any given binary.
Given this assumption, we show that at least $28\%$ (90\% credibility) of the LIGO--Virgo--KAGRA binaries contain a primary with significant spin, possibly indicative of mass-ratio reversal.
We find no evidence for binaries that contain a secondary with significant spin.
However, the single-spin framework is moderately disfavoured (natural log Bayes factor $\ln {\cal B} = 3.1$) when compared to a model that allows both black holes to spin.
If future studies can firmly establish that most merging binaries contain \textit{two} spinning black holes, it may call into question our understanding of formation mechanisms for binary black holes or the efficiency of angular momentum transport in black hole progenitors.
\end{abstract}
 
\keywords{Black holes (162) --- Compact objects (288) --- Gravitational wave astronomy (675) --- Gravitational waves (678)}

\section{Introduction} \label{sec:intro}
Recent works have suggested that angular momentum transport in black-hole progenitors may be highly efficient, leading to slowly rotating stellar cores \citep{Fuller_2019, Ma_2019}.
As a result, their eventual core collapse should produce black holes with negligible dimensionless spin magnitudes $\chi \lesssim 0.01$ \citep{Fuller_2019, Ma_2019}.
However, studies of the merging binary black hole (BBH) population observed via gravitational waves have shown that $\gtrsim 40\%$ of systems contain at least one black hole with non-negligible spin \citep{Callister_2022, Mould_2022, Tong_2022, Biscoveanu_2021, Galaudage_2021, Kimball_2021, Roulet}.

Tidal spin up is a popular explanation for the non-negligible spin observed in binary black holes \citep{Ma_2023, Fuller_2022, Hu_2022, Olejak_2021, Bavera_2020, Belczynski_2020, Qin_2018}.
This scenario begins with an isolated binary, consisting of a black hole and a companion star which is the progenitor of the second-born black hole.
The stellar companion's envelope has been stripped through binary interactions, leaving behind a Wolf-Rayet star -- a bare stellar core without an outer hydrogen envelope.
The first-born black hole induces tides on the Wolf-Rayet star that dissipate and produce a torque \citep{Ma_2023, Fuller_2022, Qin_2018, Kushnir_2017}.
As no outer layers remain to carry away this angular momentum, the rotation is retained after core collapse.

The result is a BBH system with a rapidly rotating second-born black hole.
Typically, the first-born black hole forms the primary (more massive) black hole.
However, if the binary undergoes mass-ratio reversal, the second-born (i.e., spinning) black hole will be the more massive component \citep{Broekgaarden_2022, Zevin_2022}.
If the black holes seen with gravitational waves form in the field and are tidally spun up, we expect that only one black hole in any given binary should have non-negligible spin.
However, population models for BBH spins to date assume that each component's spin is distributed independently relative to its companion's \citep[see, for example][]{GWTC3_rnp, Mould_2022}.

In this work, we model the population of merging BBH systems using a ``single-spin'' framework in which only one component in any given binary has non-negligible spin.
In doing so, we aim to ascertain whether tidal spin-up provides a good explanation for the spin properties of binary black holes observed in gravitational waves.
Within the single-spin framework, we seek to measure the fraction of mass-ratio reversed events with a spinning primary.
The fraction of mass-ratio reversed mergers can vary significantly $0-80\%$ for different models \citep{Broekgaarden_2022, Zevin_2022}.
Measuring the fraction of mass-ratio reversed mergers may therefore be useful for constraining binary evolution models.

The remainder of this work is structured as follows.
We outline our population model and inference techniques in Section~\ref{sec:method}.
In Section~\ref{sec:results} we show the results of this analysis.
We discuss the implications in Section~\ref{sec:discussion}.

\section{Method} \label{sec:method}
We propose a spin model for the BBH population that builds on previous work from \cite{Galaudage_2021} and \cite{Tong_2022}.
It is a nested mixture model that allows for three sub-populations: binaries where neither black hole spins, binaries where the primary $i=1$ black hole spins (but not the secondary $i=2$), and binaries where the secondary spins (but not the primary).
We refer to these three sub-populations as ``non-spinning,'' ``primary-spinning,'' and ``secondary-spinning.''
Note that in this framework, ``non-spinning'' is used as a proxy for a negligibly small spin $\chi_i \lesssim 0.01$ \citep[indistinguishable from $\chi_i=0$ with current measurement uncertainties; see][]{GWTC3}.
This model does not allow for the primary and secondary black holes to both spin, but we return to this possibility below using a separate model.
We assume the distribution of the two spin magnitudes $\chi_1$ and $\chi_2$ is
\begin{widetext}
\begin{equation} \label{eq:magnitude}
    \pi(\chi_1, \chi_2 | \lambda_0, \lambda_1, \mu_\chi, \sigma_\chi^2) =
    \lambda_0 \delta(\chi_1) \delta(\chi_2) + (1 - \lambda_0)\Big(
    \lambda_1 \mathrm{Beta}(\chi_1|\mu_\chi, \sigma_\chi^2) \delta(\chi_2) +
    (1 - \lambda_1) \delta(\chi_1) \mathrm{Beta}(\chi_2|\mu_\chi, \sigma_\chi^2)
    \Big).
\end{equation}
\end{widetext}
Here, $\delta(\chi_i)$ denotes the Dirac delta function, indicating a spin magnitude of zero.
Following \cite{Wysocki}, the non-zero spins are distributed according to a beta distribution with mean $\mu_\chi$ and variance $\sigma_\chi^2$.
This model assumes the $\chi_1>0$ sub-population is identical to the $\chi_2>0$ sub-population; the beta distributions for $\chi_1$ and $\chi_2$ have identical means and variances.
One can allow these two distributions to be distinct, but we find that our results do not vary meaningfully if we allow for this possibility.
The parameter $\lambda_0$ is the fraction of BBH systems with two non-spinning black holes.
Within the remaining fraction $(1 - \lambda_0)$, $\lambda_1$ is the fraction with a primary-spinning black hole as opposed to a secondary.

Following \cite{spin}, we model the (cosine) spin tilts $\cos t_i$ such that they are independently and identically distributed according to:
\begin{equation}\label{eq:orientation}
    \pi(\cos t_i|\sigma_t) =
    \mathcal{N}(\cos t_i | 1, \sigma_t)
    \Theta (\cos t_i + 1)
    \Theta (\cos t_i - 1),
\end{equation}
where $\mathcal{N}$ is a normal distribution with a mean of 1 and width $\sigma_t$, and $\Theta$ is a Heaviside step function---truncating the distribution to lie between $\cos t_i \in [-1,1]$.
This model is congruent with the assumption that spinning systems are tidally spun up, and thus should have preferentially aligned spins \citep[e.g.][]{Ma_2023}.\footnote{
Further tests that allow the mean of the spin tilt distributions to vary, allow $\cos t_1$ and $\cos t_2$ to be distributed independently, or allow for an isotropic sub-population \citep[as per][]{spin} suggest that our conclusions do not depend strongly on the model for spin orientation \citep[see also][]{Tong_2022, Vitale_2022, Mould_2022}}.
We simultaneously fit the mass and redshift distributions  using the \textsc{Power-Law + Peak} mass model from \cite{Talbot_2018} and the \textsc{Power-Law} redshift model from \cite{Fishbach_2018}.

We account for mass and redshift-based selection effects \citep[see][]{Messenger_2013, Thrane_2019, GWTC1_rnp, GWTC2_rnp, GWTC3_rnp} using the injection set from \cite{injections} \citep[see][]{Tiwari_2018, Farr_2019, Mandel_2019}.
However, we do not include spin-based effects due to sampling issues that arise at values of $\chi_i \approx 0$.\footnote{Significant code development is required to implement spin-based selection effects.}
These spin-based selection effects are believed to be relatively small for populations with less than $\sim 100$ events \citep{Ng_2018, GWTC3_rnp}.
Furthermore, these effects manifest as a bias away from effective inspiral spins $\chi_\mathrm{eff} < 0$, as well as larger uncertainties on the spin properties of $\chi_\mathrm{eff} < 0$ systems \citep{Ng_2018}.
We do not expect any correlations between such systems and the tendency to be primary or secondary spinning, thus do not expect these effects to significantly bias our results.
To test this, we draw $10^5$ events from our model, inject the corresponding signals into simulated design-sensitivity LIGO noise, and find the fraction of injections that are recovered with an optimal network signal-to-noise ratio $> 11$.
We carry out this calculation four times: assuming only primary spin, assuming only secondary spin, assuming both spin, and assuming no-spin populations.
We find that the fraction of above-detection-threshold events varies by $\lesssim 0.1 \%$ between each sub-population.
This supports our expectation that the results will not change significantly when we include spin-based selection effects.

We perform hierarchical Bayesian inference in order to measure the population hyper-parameters using gravitational-wave data from the LIGO-Virgo-KAGRA collaboration \citep[LVK;][]{LIGO, Virgo, KAGRA}.\footnote{
For a review on parameter estimation and hierarchical inference in gravitational wave astronomy, we point the reader to \cite{Thrane_2019}.}
We do so using the nested sampler \texttt{DYNESTY} \citep{dynesty} inside of the \texttt{GWPopulation} \citep{gwpopulation} package, which itself is built on top of \texttt{Bilby} \citep{bilby, bilby2}.
Our dataset begins with the 69 BBH observations from the third LVK gravitational-wave transient catalog GWTC-3 \citep{GWTC3} that were considered reliable for population analyses \citep[events with a false alarm rate $< 1 \mathrm{yr}^{-1}$;][]{GWTC3_rnp}.
However, we omit two events, GW191109\_010717 and GW200129\_065458, due to concerns related to data quality \citep{Davis_2022, Macas_2023, Payne_2022, hui}, so that we analyze 67 BBH events.
We find that the inclusion of these two events does not drastically change our results (see Section~\ref{sec:results}).

For each BBH event, we perform three sets of parameter estimation to be used in our hierarchical inference: once with a no-spin prior $\chi_1=\chi_2=0$, once with a primary-spin prior $\chi_2=0$, and once with a secondary-spin prior $\chi_1=0$.
Whichever black hole is allowed to spin is sampled with a prior that is uniform in $\chi_i$.
We use the \texttt{IMRPhenomXPHM} waveform model \citep{Pratten_2021}.
Carrying out three suites of parameter estimation runs allows us to avoid potential issues of under-sampling the posterior distribution near $\chi_i=0$ during hierarchical inference \citep[see Appendix~\ref{sec:hierarchical}, as well as][for more details]{Galaudage_2021, Tong_2022, Adamcewicz_2023}.

In order to compare the single-spin hypothesis to the hypothesis that both black holes may spin, we also construct and fit a ``both-spin'' population model.
This consists of the \textsc{Extended} model for spin magnitude from \cite{Tong_2022} (see their Eq.~2), combined with our simplified model for spin orientation defined in Eq.~\ref{eq:orientation}.
We obtain a fourth set of parameter estimation results in which both black holes may have non-zero spins in order to perform hierarchical inference with this both-spin population model.

We set uniform priors over $[0,1]$ for the mixing fractions $\lambda_0$ and $\lambda_1$.
The priors on other population hyper-parameters are identical to those used in \cite{Tong_2022}.
These priors do not allow for singularities in the $\chi_i$ Beta distributions.

\section{Results} \label{sec:results}
First, we compare the evidence for the ``single-spin'' population model proposed in Section~\ref{sec:method} to the previously used ``both-spin'' model in which both black holes in any given binary may spin.
We find that the single-spin model is disfavoured by a natural log Bayes factor of $\ln \mathcal{B} = 3.1$ (difference in maximum natural log likelihood of $\Delta \ln \mathcal{L}_{\max} = 2.0$).
The single-spin model incurs an Occam penalty for its added complexity relative to the both-spin model, which does not have the $\lambda_1$ parameter.
However, the both-spin model also yields a better overall fit, evident by its larger maximum likelihood.
The numerical value of $\ln \mathcal{B} = 3.1$ is not large enough to draw a strong conclusion that the both-spin model is clearly preferred over the single-spin model.\footnote{
We find that we can draw the same conclusion when including single-spin and both-spin binaries in the same population model: a dominant sub-population of secondary-spin black holes is ruled out, while both-spin binaries are modestly preferred over primary-spin only systems.
Due to an Occam penalty, this model is disfavoured when compared to the simpler both-spin model by a natural log Bayes factor $\ln \mathcal{B} = 1.2$.
No noteworthy covariance in mixing fractions arises within this framework.
}
It is, however, an interesting preliminary result that we are keen to revisit as more data becomes available.
When including GW191109\_010717 and GW200129\_065458 (or either event on its own), we find that support for the both-spin model increases by $\Delta \ln \mathcal{B} \lesssim 1$.

\begin{figure}
    \centering
    \includegraphics[width=\columnwidth]{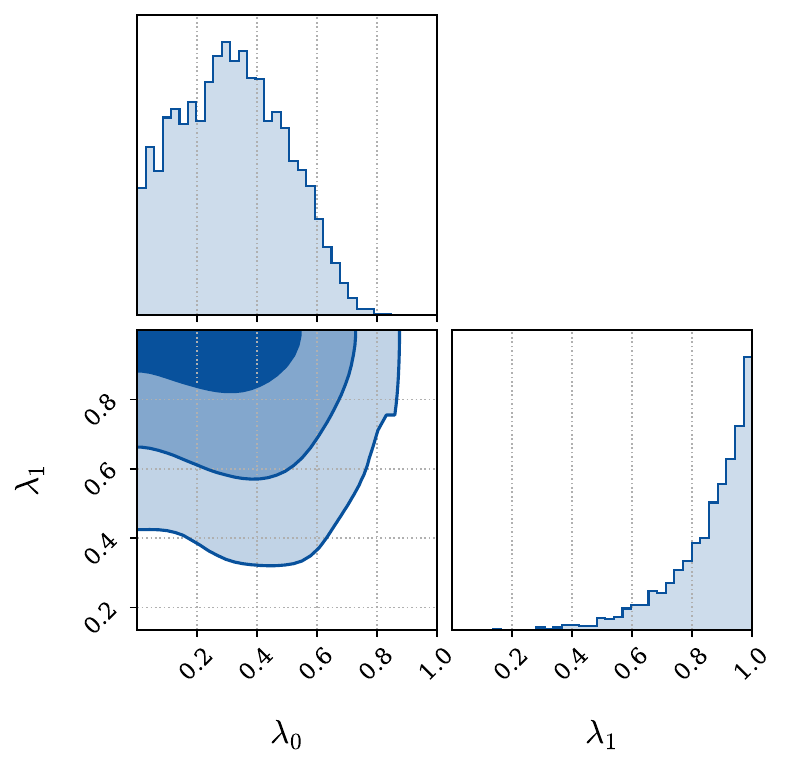}
    \caption{
    Posterior corner plot for the fraction of BBH systems with negligible spin $\lambda_0$, and the fraction of spinning BBH systems with $\chi_1>0$, $\lambda_1$.
    The different shades indicate the 50\%, 90\%, and 99\% credible intervals.
    The fact that $\lambda_0=1$ is ruled out is already well-established: at least some binary black hole systems contain a black hole with non-negligible spin.
    The fact that $\lambda_1=0$ is ruled out suggests that---within the single-spin framework, and among binaries with a spinning black hole---it is the primary mass black hole that is spinning at least $59\%$ of the time.
    Under the assumption of single-spin, the data are consistent with the possibility that only the primary black hole spins.
    }
    \label{fig:frac_corner}
\end{figure}

Next, we set aside for a moment the possibility that both black holes have non-negligible spin, and assume that spinning black holes are tidally spun up (i.e., that there can be at most one black hole with non-negligible spin in each binary).
In Fig.~\ref{fig:frac_corner}, we show the posterior corner plot for the mixing fractions $\lambda_0$ (the fraction of events with non-negligible spin) and $\lambda_1$ (the fraction of spinning events with $\chi_1>0$ as opposed to $\chi_2>0$).
We show posterior distributions for other population hyper-parameters governing BBH spins in Appendix~\ref{sec:appendix}.
With 90\% credibility, we measure the fraction of non-spinning systems to be $\lambda_0 \leq 0.60$ -- consistent with the results of \cite{Tong_2022}.
Amongst BBH systems with measurable component spins, we find the fraction of primary-spinning systems to be $\lambda_1 \geq 0.59$ (90\% credibility).
We rule out $\lambda_1=0$ with high credibility.
The posterior is peaked at $\lambda_1=1$, the point in parameter space where no secondary black holes have appreciable spin.
These results do not change meaningfully when including GW191109\_010717, GW200129\_065458, or both.

As a check, we analyze 10 simulated signals with $\chi_1=0$ and $\chi_2>0$ drawn from a beta distribution.
As expected, the resulting posterior from hierarchical inference peaks at $(\lambda_0=0, \lambda_1=0$), which assures us that our result does not arise from some pathological prior effect.

\begin{figure*}
    \centering
    \includegraphics[width=1.5\columnwidth]{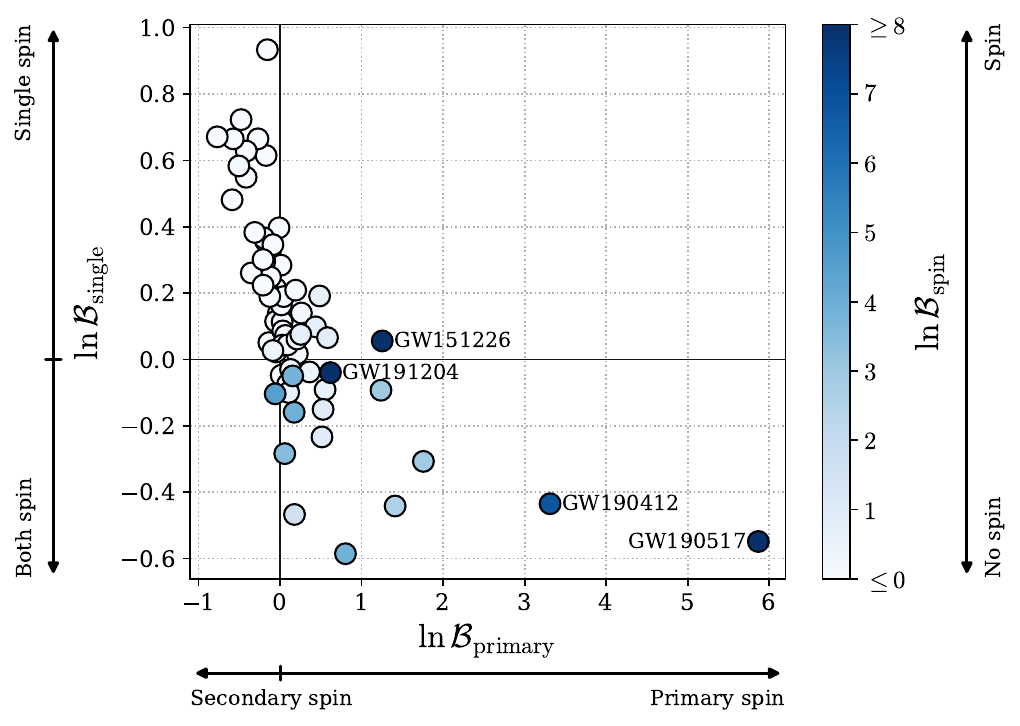}
    \caption{A comparison of each spin hypothesis for each event.
    Each point represents a single BBH event observed in gravitational waves.
    The colour of the point represents the natural log Bayes factor for the spinning hypothesis $\ln \mathcal{B}_\mathrm{spin}$; white circles are events best explained as non-spinning while blue circles are best explained as containing at least one spinning black hole.
    The vertical axis shows the natural log Bayes factor comparing the single-spin hypothesis (positive values) to the both-spin hypothesis (negative values) $\ln \mathcal{B}_\mathrm{single}$. 
    The horizontal axis shows the natural log Bayes factor comparing the primary-spin hypothesis (positive values) to the secondary-spin hypothesis (negative values) $\ln \mathcal{B}_\mathrm{primary}$. 
    Note there is significantly less spread in ${\cal B}_\mathrm{single}$ than there is in ${\cal B}_\mathrm{primary}$, indicating that it is relatively difficult to ascertain if the primary is spinning by itself.
    The four events with the highest evidence for spin (darkest blue) are labelled, being GW190412\_053044, GW190517\_055101, GW151226, and GW191204\_171526.
    }
    \label{fig:evidences}
\end{figure*}

It is useful to understand which features in the data are most responsible for our results.
In Fig.~\ref{fig:evidences}, we plot the evidence obtained during the initial parameter estimation for each event, given each different spin hypothesis (see Section~\ref{sec:method} and Appendix~\ref{sec:hierarchical} for how these evidence values are used in the hierarchical analysis presented above).
In this scatter plot, the horizontal axis is the natural log Bayes factor comparing the primary-spin evidence ${\cal Z}_1$ to the secondary-spin evidence ${\cal Z}_2$:
\begin{equation}
    \ln \mathcal{B}_\mathrm{primary} =
    \ln \left(\frac{\mathcal{Z}_1}{
    \mathcal{Z}_2}\right).
\end{equation}
The vertical axis is the natural log Bayes factor comparing the single-spin evidence (${\cal Z}_1$ or ${\cal Z}_2$---whichever is larger) to the both-spinning evidence ${\cal Z}_b$:
\begin{equation}
    \ln \mathcal{B}_\mathrm{single} =
    \ln \left(\frac{
    \max(\mathcal{Z}_1, \mathcal{Z}_2)}{
    \mathcal{Z}_b}\right).
\end{equation}
Meanwhile, the color bar shows the natural log Bayes factor comparing the spinning hypothesis (whichever is largest) to the no-spin hypothesis:
\begin{equation}
    \ln \mathcal{B}_\mathrm{spin} =
    \ln \left(\frac{
    \max(\mathcal{Z}_1, \mathcal{Z}_2, \mathcal{Z}_b)}{
    \mathcal{Z}_0}\right).
\end{equation}
Events with evidence for spin (blue dots) show a small preference to lie below zero in the vertical axis, indicating a preference for the both-spin hypothesis.
This amalgamates as a moderate preference for both-spin systems over single-spin systems on a population level ($\ln \mathcal{B} = 3.1$ from above).
These events tend to show a much larger deviation from zero in the horizontal axis -- trending towards values of $\ln {\cal B}_\mathrm{primary}>0$, indicating a preference for primary-spin as opposed to secondary-spin.
The events that prefer secondary-spin over primary-spin ($\ln {\cal B}_\mathrm{primary}<0$) tend to be white circles, indicating that these events are best explained as not spinning at all.
Under the assumption of single-spin, this results in a strong preference for primary-spin systems on the population-level, as seen in Fig.~\ref{fig:frac_corner}.

\section{Discussion} \label{sec:discussion}
We find that the spin properties of the binary black holes in GWTC-3 require at least $28\%$ of all binaries to include a primary with non-negligible spin.
Among spinning binaries, we find that at least $59\%$ of systems have a non-negligible primary spin.
These binaries are either mass-ratio reversed or contain \textit{two} black holes with non-negligible spin.

This result is consistent with predictions from \cite{Zevin_2022} and \cite{Broekgaarden_2022}, which suggest that up to $\approx 72 - 82 \%$, of the BBH population may be mass-ratio reversed.
Furthermore, \cite{Farah_2023} find that a large fraction of the BBH population is consistent with mass-ratio reversal due to asymmetries in the distributions of primary and secondary masses.
While our posterior for the fraction of primary spinning systems $\lambda_1$ is consistent with $0.82$, it peaks at $\lambda_1=1$.
With additional events, it may be possible in the near future to distinguish between $\lambda_1=0.80$ and $\lambda_1=1$.
A strong preference for $\lambda_1=1$ would be difficult to explain within the standard field formation scenario given our current understanding of angular momentum transport and tidal spin-up \citep[see][for example]{Qin_2018, Fuller_2019, Ma_2023}.
Of course, if a strong statistical preference for the both-spinning framework can be established, then the entire discussion of mass-ratio reversal may be moot.
It would be difficult to explain such a result within the field binary framework, unless angular momentum transport in massive stars is less efficient than expected \citep[see][and discussions therein]{Heger_2005, Qin_2018, Fuller_2019}.
On this point, \cite{Callister_2021} find that if spinning binary black holes are formed in the field and undergo tidal spin up, extreme natal kicks are required to produce the observed range of spin tilts in the LVK data.
\cite{Callister_2021} suggest that inefficient angular momentum transport in black hole progenitors (thus non-negligible spins for first-born black holes) may alleviate this requirement for extreme kicks.
This is because the first-born black hole forms when the binary has a greater orbital separation so is more easily misaligned by smaller natal kicks \citep{Callister_2021}.
However, these findings are disputed by \cite{Stevenson2022}, who question the assumption that all secondary-mass black holes can be tidally spun up.

\cite{Qin_2022} highlights the BBH merger GW190403\_051519: an event that does not pass the threshold for inclusion in population studies, yet has the highest inferred effective inspiral spin of any LVK observation to date $\chi_\mathrm{eff} \approx 0.7$.
This event, if authentic, provides a strong signature for a rapidly spinning primary $\chi_1 = 0.92^{+0.07}_{-0.22}$, making it consistent with the results presented here.

The event GW190412\_53044 shows the second strongest evidence for a primary spin component in our analyses.
Motivated by the tidal spin up hypothesis, \cite{Mandel_2020} suggest that GW190412\_53044 can be explained as a system with a rapidly rotating secondary by imposing a prior in which the primary is assumed to have negligible spin.
However, \cite{Zevin_2020} argue that this assumption is statistically disfavoured by the data---a point that we reiterate in Fig.~\ref{fig:evidences}.

We come to a similar conclusion as \cite{Mould_2022}, in that both studies suggest that both black holes spinning in any given binary is the best description for the majority of the population.
\cite{Mould_2022} suggest that such systems make up $\approx 77\%$ of binary black holes.
However, \cite{Mould_2022} find that a larger fraction of BBH systems may be described as secondary-spinning ($\lesssim 42\%$) as opposed to primary-spinning ($\lesssim 32\%$).
In contrast, setting aside the possibility that both black holes may spin, our results suggest that a much larger fraction of BBH systems can be described as primary-spinning ($\lesssim 88\%$) as opposed to secondary-spinning ($\lesssim 28\%$).
While our inferences on these fractions may decrease when a both-spin sub-population is factored in, the \textit{relative} proportion of primary-spin to secondary-spin systems does not vary meaningfully.
Furthermore, the results of \cite{Mould_2022} suggest that $\lesssim 6\%$ of binaries can be described with both black holes having negligible spins, while we measure this fraction to be $\lesssim 60 \%$.
Our results, however, are consistent with the findings of \cite{Tong_2022}, \cite{Callister_2022}, and \cite{Roulet} on this front.
These discrepancies may be due to a number of factors.
Firstly, our spin models are set up to explicitly test the hypothesis that only one black hole spins in any given binary whereas \cite{Mould_2022} endeavor to measure the distributions of $\chi_1$ and $\chi_2$ \textit{independently}, without attempting to force one or more spin magnitudes to zero in each binary.
Also, \cite{Mould_2022} do not use dedicated samples near $\chi_i \approx 0$, which may lead to issues of under-sampling this region.

Another possibility is that a large fraction of the events in GWTC-3 ($\gtrsim 28\%$) are not formed in the field \citep{Zevin2021}.
Binaries merging in dense stellar environments can merge repeatedly through hierarchical mergers \citep[e.g.][]{Fishbach_2017, Gerosa_2017, Rodriguez_2019, Doctor_2020, Doctor_2021}.
While many analyses suggest that the LVK data is consistent with isotropy in BBH spin tilts \citep{GWTC3_rnp, Vitale_2022, Callister_2022, callister_2023, golomb_2023, Edelman_2023}, other studies have found that binary black hole spin may tend towards alignment with the orbital angular momentum \citep[e.g.][]{Roulet,Galaudage_2021,Tong_2022}.
If it is true that BBH spins are preferentially aligned, this would conflict with what one would expect for binaries formed in globular clusters \citep{Rodriguez_2016,Farr,Stevenson,spin,Vitale_2017,Yu_2020}.

Active galactic nuclei (AGN) may provide an environment in which both black holes can be spun up (through both accretion and hierarchical mergers) while providing a preferred axis with which to align black hole spin \citep{Bogdanovic_2007, agn_avi, mckernan_2023}.
Namely, prograde accretion of gas in AGN disks may simultaneously spin up merging black holes (primary or secondary) and torque them into alignment with the disk's rotation \citep{Bogdanovic_2007, mckernan_2023}.
At the same time, the high stellar densities and escape velocities of AGNs provide a suitable environment for potential hierarchical mergers, producing black holes with $\chi_i \approx 0.7$ \citep[see, for example][]{Tichy_2008}---and large masses like the components of GW190521 \citep{GW190521,GW190521_implications}.
These black holes may then be spun down to magnitudes consistent with gravitational-wave observations ($\chi_i \approx 0.2-0.4$) via retrograde accretion of gas in the AGN disk \citep{mckernan_2023}.

Black holes observed in high-mass X-ray binaries (with lower-mass companions) appear to have large aligned spins $\chi_1 \gtrsim 0.8$ \citep{Liu_2008, Miller_Jones_2021, Reynolds_2021}, often thought to be a result of accretion \citep{Podsiadlowski_2003, Qin_2019, Shao_2020}.\footnote{
Although, these measurements may be affected by systematic uncertainties, resulting in errors on the order of $\chi_1 \sim 0.1$ \citep{Taylor_2018, Salvesen_2020, Falanga_2021}.}
If primary-spin systems are common in gravitational waves, this may indicate that merging binary black holes can undergo a similar evolutionary process to high-mass X-ray binaries \citep{Fishbach_2022, Gallegos_Garcia_2022, Shao_2022}.
Assuming that case-A mass transfer is responsible for spinning up the primary, \cite{Gallegos_Garcia_2022} find that up to $\approx 20 \%$ of BBH mergers may be former high-mass X-ray binaries.
However, previous studies show that the large spin magnitudes of black holes in high-mass X-ray binaries are in tension with the BBH spin distribution from gravitational waves \citep{Roulet_2019, Fishbach_2022}.
This tension may be somewhat relieved when modeling the BBH population under the assumption of rapidly-spinning primaries \citep{Fishbach_2022}.
On this note, we find increased support for large spin magnitudes when using the single-spin framework (see Fig.~\ref{fig:extra_corner} from Appendix~\ref{sec:appendix}).
We find that the single-spin framework allows for up to $\approx 10\%$ of binary black holes to have $\chi_i \geq 0.8$, while the (preferred) both-spin model suggests these systems may only make up $\lesssim 3\%$ of the population (90\% credibility).
This is most likely driven by support for high values of $\chi_\mathrm{eff}$ in the data (predominantly from the dark blue events in Fig.~\ref{fig:evidences}).
When only one black hole is allowed to spin, higher spin magnitudes are required to reach these large values of $\chi_\mathrm{eff}$ when compared to the scenario in which both black holes may spin.

\section*{Acknowledgements}
We thank our anonymous reviewer, Amanda Farah, Christopher P. L. Berry, Sylvia Biscoveanu, Thomas Callister, Maya Fishbach, Matthew Mould, and Salvatore Vitale for helpful comments on this manuscript.
We acknowledge support from the Australian Research Council (ARC) Centre of Excellence CE170100004, LE210100002, and ARC DP230103088.
This material is based upon work supported by NSF's LIGO Laboratory which is a major facility fully funded by the National Science Foundation.
The authors are grateful for computational resources provided by the LIGO Laboratory and supported by National Science Foundation Grants PHY-0757058 and PHY-0823459.

This research has made use of data or software obtained from the Gravitational Wave Open Science Center (gw-openscience.org), a service of LIGO Laboratory, the LIGO Scientific Collaboration, the Virgo Collaboration, and KAGRA. LIGO Laboratory and Advanced LIGO are funded by the United States National Science Foundation (NSF) as well as the Science and Technology Facilities Council (STFC) of the United Kingdom, the Max-Planck-Society (MPS), and the State of Niedersachsen/Germany for support of the construction of Advanced LIGO and construction and operation of the GEO600 detector. Additional support for Advanced LIGO was provided by the Australian Research Council. Virgo is funded, through the European Gravitational Observatory (EGO), by the French Centre National de Recherche Scientifique (CNRS), the Italian Istituto Nazionale di Fisica Nucleare (INFN) and the Dutch Nikhef, with contributions by institutions from Belgium, Germany, Greece, Hungary, Ireland, Japan, Monaco, Poland, Portugal, Spain. The construction and operation of KAGRA are funded by Ministry of Education, Culture, Sports, Science and Technology (MEXT), and Japan Society for the Promotion of Science (JSPS), National Research Foundation (NRF) and Ministry of Science and ICT (MSIT) in Korea, Academia Sinica (AS) and the Ministry of Science and Technology (MoST) in Taiwan.

\appendix

\section{Population likelihoods with sharp features}
\label{sec:hierarchical}
When introducing sharp features (like a non-spinning peak) into a population model, one can encounter issues of under-sampling during hierarchical inference.
That is to say, an insufficient number of fiducial (event-level) samples in the relevant region of parameter space makes the sharp feature in the population model difficult to resolve via re-weighting.
This may lead to spurious inferences.

To counteract this, we use the method presented in \cite{Galaudage_2021}, \cite{Tong_2022}, and \cite{Adamcewicz_2023}.
Each variation of the zero-spin peak (no-spin $\chi_1=\chi_2=0$, primary-spin $\chi_2=0$, and secondary-spin $\chi_1=0$; see Section~\ref{sec:method}) is imposed on the prior during separate event-level parameter estimation analyses (reiterating, wherever $\chi_i>0$, the prior is uniform).
In effect, this means that there are an abundance of samples within each variation of the zero-spin peak, and spin samples only require re-weighting outside of these sharp features (when $\chi_i>0$).

Using these three sets of parameter estimation results, we can construct a population likelihood following the model from Section~\ref{sec:method} as
\begin{equation} \label{eq:pop_likelihood}
    \mathcal{L}(d|\Lambda, \lambda_0, \lambda_1) =
    \prod_i^N \frac{1}{n} \left[
    \lambda_0 \mathcal{Z}_0^i \sum_k^n w_0(\theta_k^i|\Lambda)
    + (1 - \lambda_0) \left(
    \lambda_1 \mathcal{Z}_1^i \sum_k^n w_1(\theta_k^i|\Lambda)
    + (1 - \lambda_1) \mathcal{Z}_2^i \sum_k^n w_2(\theta_k^i|\Lambda)
    \right)
    \right].
\end{equation}
Here, $\lambda_0$ and $\lambda_1$ are the mixing fractions for different spin configurations (see Section~\ref{sec:method}), $\Lambda$ denotes the set of all other population hyper-parameters (e.g. $\mu_\chi$ and $\sigma_\chi^2$), $N$ is the number of events, and $n$ is the total number of samples per-event obtained from parameter estimation.
The Bayesian evidence obtained during parameter estimation for event $i$, with spin configuration $s$ (no-spin $s=0$, primary-spin $s=1$, and secondary-spin $s=2$) is denoted $\mathcal{Z}_s^i$.
Finally, the weights $w_s$ for a given parameter estimation sample $\theta_k^i$ are
\begin{equation}
    w_s(\theta|\Lambda) =
    \frac{\pi_s(\chi_1,\chi_2,\cos t_1,\cos t_2|\Lambda)\pi(m_1, q, z|\Lambda)}
    {\pi(\chi_1,\chi_2,\cos t_1, \cos t_2, m_1, q, z|\text{\O})},
\end{equation}
where $\pi(\theta|\text{\O})$ is the prior probability from parameter estimation, $\pi(m_1, q, z|\Lambda)$ is the population model for primary mass, mass ratio, and redshift from Section~\ref{sec:method}, and
\begin{align}
    \pi_0(\chi_1,\chi_2,\cos t_1,\cos t_2|\Lambda) &=
    \delta(\chi_1)\delta(\chi_2), \\
    \pi_1(\chi_1,\chi_2,\cos t_1,\cos t_2|\Lambda) &=
    \mathrm{Beta}(\chi_1|\mu_\chi,\sigma_\chi^2)\delta(\chi_2)\pi(\cos t_1|\sigma_t), \\
    \pi_2(\chi_1,\chi_2,\cos t_1,\cos t_2|\Lambda) &=
    \delta(\chi_1)\mathrm{Beta}(\chi_2|\mu_\chi,\sigma_\chi^2)\pi(\cos t_2|\sigma_t),
\end{align}
gives the spin model for each configuration.
Again, note that only the uniformly sampled $\chi_i>0$ spin values are re-weighted (to follow a Beta distribution).
Also note that we have omitted the detection efficiency factor that accounts for selection effects from Eq.~\ref{eq:pop_likelihood} for the sake of brevity.

\section{Additional plots} \label{sec:appendix}
\begin{figure}
    \centering
    \includegraphics[width=0.6\columnwidth]{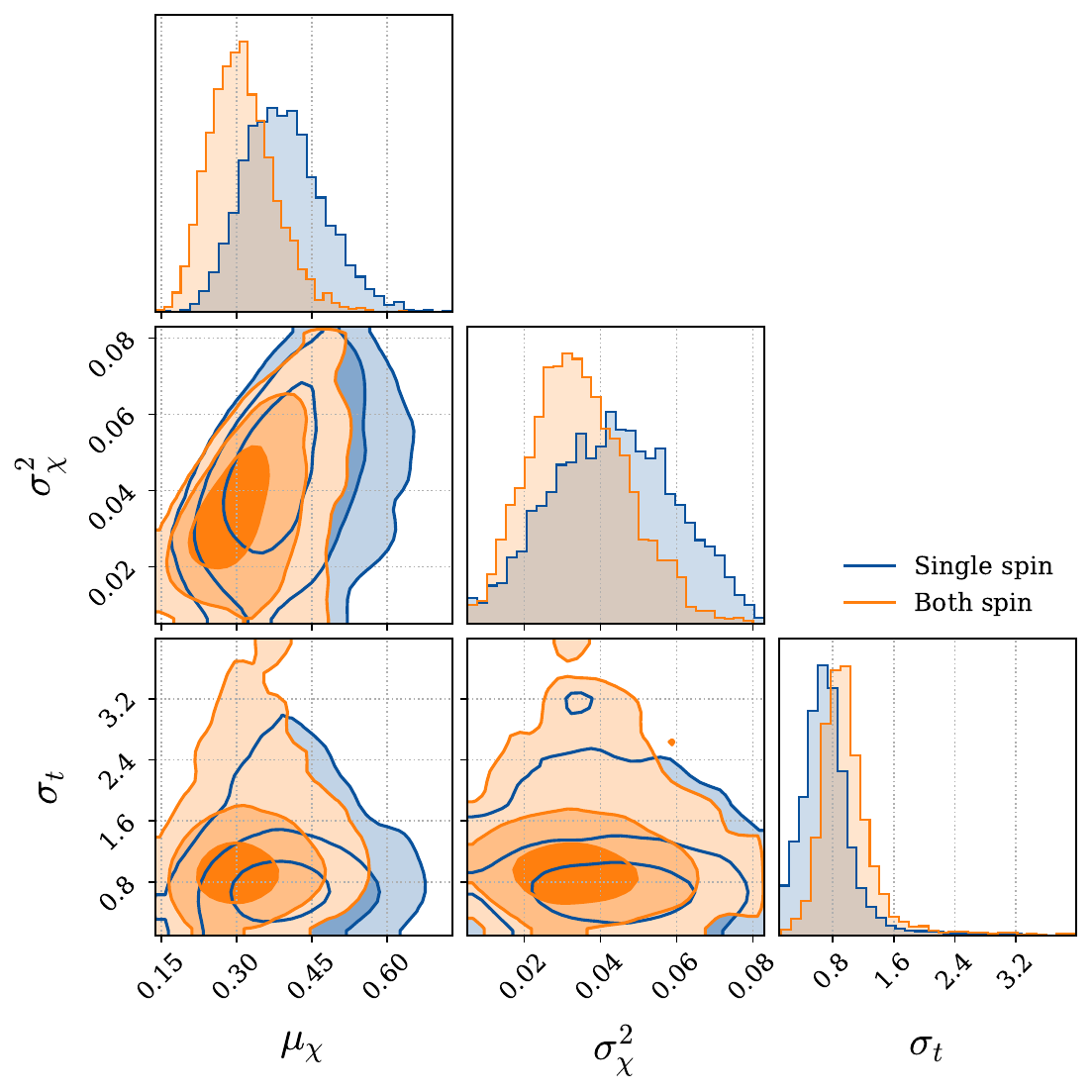}
    \caption{Posterior corner plot for hyper-parameters governing the BBH spin distribution.
    Blue shows the posteriors given the single-spin model (Eq.~\ref{eq:magnitude}), while orange shows the posteriors given the both-spin model \citep[Eq.~2 from][]{Tong_2022}.
    Both models use Eq.~\ref{eq:orientation} for spin orientation.
    Contours on the two-dimensional plot give the 50\%, 90\% and 99\% credible regions.
    In the single-spin framework, we measure the mean and variance of the spin magnitude distribution to be $\mu_\chi = 0.39^{+0.14}_{-0.11}$ and $\sigma_\chi^2 = 0.04^{+0.03}_{-0.03}$ respectively.
    Meanwhile, the width of the cosine spin tilt distribution is found to be $\sigma_t = 0.70^{+0.64}_{-0.44}$.
    In the both-spin framework, we find the mean and variance of the spin magnitude distribution to be $\mu_\chi = 0.31^{+0.12}_{-0.09}$ and $\sigma_\chi^2 = 0.03^{+0.02}_{-0.02}$, with the width of the cosine spin tilt distribution being $\sigma_t = 0.93^{+0.73}_{-0.39}$.}
    \label{fig:extra_corner}
\end{figure}

\begin{figure}
    \centering
    \includegraphics[width=0.8\columnwidth]{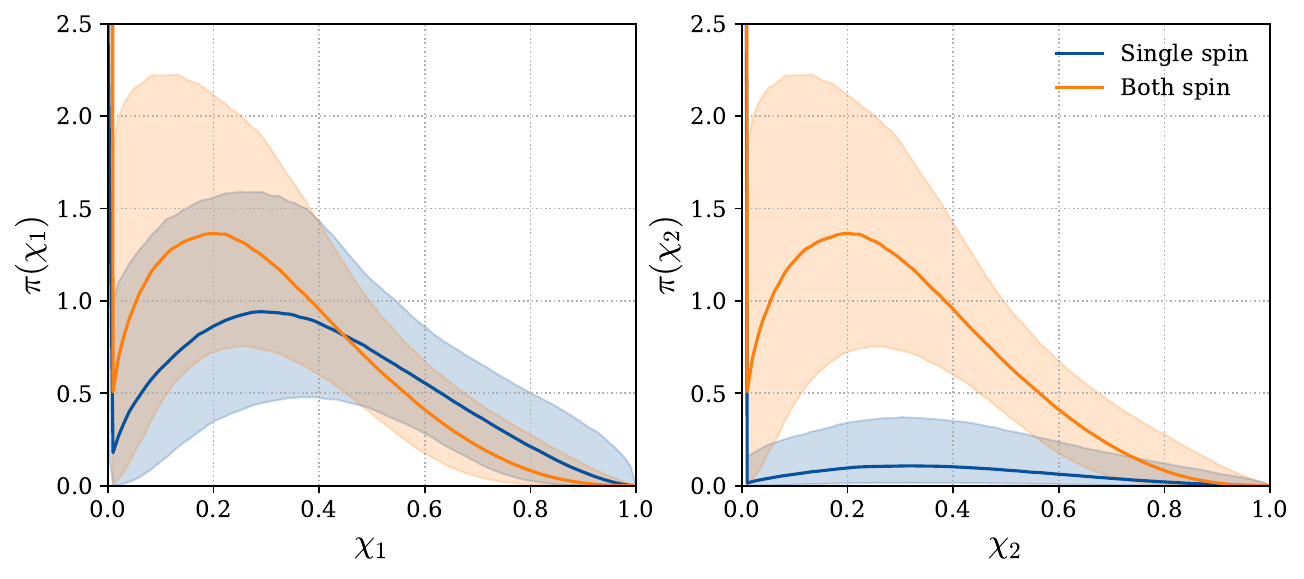}
    \caption{Population predictive distributions for primary (left) and secondary (right) spin magnitude.
    Blue shows the single-spin model (Eq.~\ref{eq:magnitude}), while orange shows the both-spin model \citep[Eq.~2 from][]{Tong_2022}.
    In either model the distributions for non-zero $\chi_1$ and $\chi_2$ must have identical shapes.
    In the both-spin model, they must also have identical amplitudes.
    The both-spin model enforces that either $\chi_1 = \chi_2 = 0$ or that $\chi_1 > 0$ and $\chi_2 > 0$ simultaneously.
    Meanwhile, the single-spin model enforces that $\chi_1 = \chi_2 = 0$, that $\chi_1 > 0$ and $\chi_2=0$, or that $\chi_1 = 0$ and $\chi_2 > 0$.}
    \label{fig:chi_pop}
\end{figure}

\begin{figure}
    \centering
    \includegraphics[width=0.5\columnwidth]{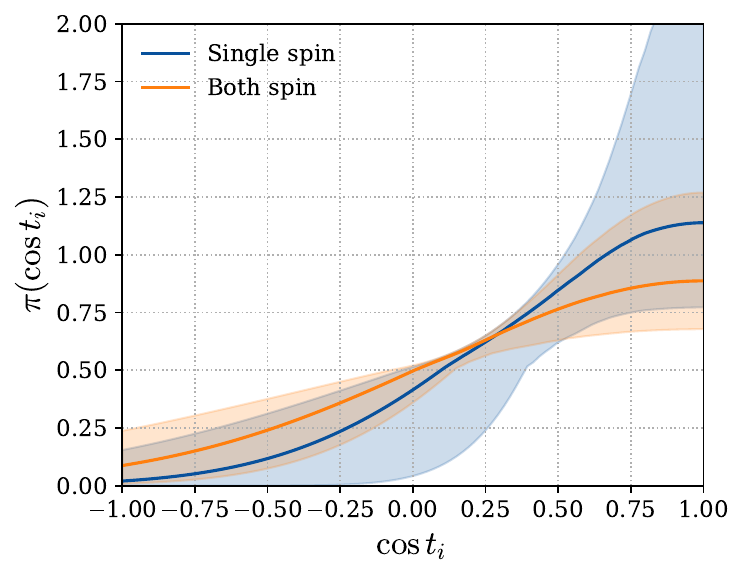}
    \caption{Population predictive distributions for cosine spin tilts following Eq.~\ref{eq:orientation}.
    Blue shows the results for the single-spin model (Eq.~\ref{eq:magnitude}), while orange shows the results for the both-spin model \citep[Eq.~2 from][]{Tong_2022}.}
    \label{fig:tilt_pop}
\end{figure}

We include an additional corner plot for the remaining population hyper-parameters governing the BBH spin distribution for the single-spin and both-spin models in Fig.~\ref{fig:extra_corner}.
We also plot population predictive distributions (for both models) for spin magnitudes and tilts in Fig.~\ref{fig:chi_pop} and Fig.~\ref{fig:tilt_pop} respectively.

\bibliographystyle{aasjournal}
\bibliography{refs}

\end{document}